# A generalizable approach based on U-Net model for automatic Intra retinal cyst segmentation in SD-OCT images

Razieh Ganjee

*The Faculty of Computer Science and Engineering, Shahid Beheshti University G.C, Tehran, Iran*

*r_ganjee@sbu.ac.ir*

Mohsen Ebrahimi Moghaddam[*]

*The Faculty of Computer Science and Engineering, Shahid Beheshti University G.C, Tehran, Iran*

*\*Corresponding author: m_moghadam@sbu.ac.ir*

Ramin Nourinia

*Ophthalmic research center, Shahid Beheshti University of Medical Sciences, Tehran, Iran*

*ramin.retin@gmail.com*

*Abstract*

Intra retinal fluids or Cysts are one of the important symptoms of macular pathologies that are efficiently visualized in OCT images. Automatic segmentation of these abnormalities has widely investigated in medical image processing studies. In this paper, we propose a new U-Net-based approach for Intra retinal cyst segmentation across different vendors that improves some of the challenges faced by previous deep-based techniques. The proposed method has two main steps: 1- prior information embedding and input data adjustment, and 2- IRC segmentation model. In the first step, we inject the information into the network in a way that overcomes some of the network limitations in receiving data and learning important contextual knowledge. And in the next step, we introduced a connection module between encoder and decoder parts of the standard U-Net architecture that transfers information more effectively from encoder to decoder part. Two public datasets namely OPTIMA and KERMANY were employed to evaluate the proposed method. Results showed that the proposed method is an efficient vendor independent approach for IRC segmentation with mean Dice values of 0.78 and 0.81 on the OPTIMA and KERMANY datasets, respectively.

*1.Introduction*

Optical coherence tomography (OCT) is a non-invasive imaging technology that currently is widely used to diagnose and monitor the progression of ocular diseases. This imaging modality is a useful tool for visualizing morphological retinal tissue variations that occur due to macular disease (Ganjee et al., 2018; Nazir et al., 2019). One of the macular pathologies that is efficiently visualized using this imaging modality is cystoid macular edema (CME) (Moschos, 2008). CME occurs due to vascular defects that lead to fluid leakage into the retina. If CME patients are not treated in time, they may lose their central vision forever. Therefore, automatic segmentation of intra-retinal cystoid fluid (IRC) or cysts in OCT images is valuable and can help ophthalmologists to assess the treatment progress efficiently. However, automatic segmentation of cysts obtained by various vendors is a challenging problem because OCT images captured by various vendors have different resolutions, intensity variation, and noise levels.

Over the past years, several automatic methods have been presented for automatic segmentation of cysts (X. Chen et al., 2012; Chiu et al., 2015; de Sisternes et al., 2015; Girish et al., 2018; González et al., 2013; Gopinath & Sivaswamy, 2016; Lee et al., 2017; Quellec et al., 2010; Rashno et al., 2018; Roy et al., 2017; Roychowdhury et al., 2013; Schlegl et al., 2018, 2015; Venhuizen, F., van Grinsven, M. J., van Ginneken, B., Hoyng, C. C., Theelen, T., Sanchez, 2016; Wang et al., 2016; Wilkins et al., 2012; Zhu et al., 2017) but few of them such as (de Sisternes et al., 2015; Girish et al., 2019; Gopinath & Sivaswamy, 2019; Oguz et al., 2016; Venhuizen, F., van Grinsven, M. J., van Ginneken, B., Hoyng, C. C., Theelen, T., Sanchez, 2016; Venhuizen et al., 2018) are vendor-independent and address this challenge. For the first time, in 2015, OPTIMA cyst segmentation challenge (*Optima cyst segmentation challenge*, 2015) published a database with a variety of retinal cysts provided by 4 different vendors consists of Cirrus, Topcon, Spectralis, and Nidek to evaluate segmentation methods in terms of vendor independency. Different approaches participated in this challenge including a simple classifier trained on the 34 hand-crafted features (de Sisternes et al., 2015), a patch-based CNN method (Venhuizen, F., van Grinsven, M. J., van Ginneken, B., Hoyng, C. C., Theelen, T., Sanchez, 2016), a rule-based method by employing center-surrounded filters and random forest classifier (Gopinath & Sivaswamy, 2016), and two unsupervised graph-based (Oguz et al., 2016) and curvelet-based (Esmaeili et al., 2016) methods. Among these methods, the rule-based method (Gopinath & Sivaswamy, 2016) could not operate as well as others on the test set and failed to segment IRC regions appropriately. The method proposed by Esmaeili et al. (Esmaeili et al., 2016) was vendor-specific and was only evaluated on Spectralis images. The only deep learning-based approach that participated in this challenge was proposed by Venhuizen et al. (Venhuizen, F., van Grinsven, M.

J., van Ginneken, B., Hoyng, C. C., Theelen, T., Sanchez, 2016). In this method, three CNNs with three different patch sizes were used to manage cysts size variety. Hence, training of this method is memory and time consuming, and also the efficiency of that is strongly dependent on the patch size selection. The winner of this challenge was the method proposed by de Sisternes et al. (de Sisternes et al., 2015). However, this method needs to segment retinal layers accurately to compute features, while middle retinal layer segmentation in the presence of cysts is yet an open challenging problem.

After the Optima challenge, development of the vendor-independent IRC segmentation method was also addressed in later works presented in (Girish et al., 2019; Gopinath & Sivaswamy, 2019; Venhuizen et al., 2018). A cascade of two U-Net architecture-based networks was employed in (Venhuizen et al., 2018). They employed the first U-Net for segmentation of retina area and then integrated the output of this network with the second U-Net for segmenting of IRC regions. Another development of the vendor-independent method based on the U-Net model was published in (Girish et al., 2019). They attempted to customize U-Net parameters such as the optimal number of layers and the optimal kernel dimensions for the IRC segmentation problem. In (Gopinath & Sivaswamy, 2019), a generalized motion pattern (GMP) and deep learning were employed for segmentation of cysts. They used a CNN to learn a function so that cysts are enhanced while other normal tissues are suppressed. Then, IRCs are extracted by clustering the probability map obtained from CNN. The most recent IRC segmentation methods were proposed in (Lu et al., 2019) and our previous paper (Ganjee et al., 2020), respectively. In Lu et al. (Lu et al., 2019) method, three types of fluids including IRC, PED (Pigment Epithelial Detachment), and SRF (subretinal fluid), were segmented by employing a large training set captured by three different types of OCT devices. In this work, a separate network based on the U-Net was trained on images from each OCT imaging device. Hence, their method is not vendor-independent. However, by employing this strategy, they could obtain the first rank in the MICCAI RETOUCH (Bogunovic et al., 2019) challenge. To the best of our knowledge, our previous unsupervised approach (Ganjee et al., 2020) is the newest vendor-independent IRC segmentation method that segments cysts by limiting the searching space toward the targets in three levels of a hierarchical framework. Although our previous unsupervised method could obtain valuable results and competitive with the deep-based methods, deep-learning-based methods are more of interest in recent years.

As it was reviewed, the previous deep learning-based methods followed two different approaches to dealing with the IRC segmentation in SD-OCT images captured by different types of OCT imaging devices; in one approach input images are asymmetrically resized to an equal size (Girish et al., 2019; Gopinath & Sivaswamy, 2019; Venhuizen et al., 2018) and in the other approach, a separate network is trained for each of the vendor (Lu et al., 2019). In the first approach, asymmetrically resizing the input images causes to loss of small cysts, merging cysts placed near each other and preventing lesions to appear with their real shape, size, and location. And the second approach is not vendor-independent because of using a separate network for each vendor. In this study, we proposed a vendor-independent approach based on the U-Net architecture that is summarized in two parts: 1- prior information embedding and input data adjustment, and 2- IRC segmentation model. In the first part, while the asymmetrically resizing problem was solved, we provided the possibility of expert knowledge injection to the network in order to manage some of the limitations of U-Net-based models in learning location of target objects and receiving data with the same size. And in the second part, by adding a connection module to the core of the U-Net architecture, between encoder and decoder parts, we both increased field of view (FOV) and changed the focus of the model on salient regions. Therefore, this connection module helps U-Net to transfer more useful and meaningful features from encoder to decoder part resulting in more accurate segmentation. In the following, first, the proposed method is explained in detail in part 2, then the evaluation results are presented in part 3, and finally, in parts 4 and 5, discussion and conclusion are discussed, respectively.

## *2.Proposed method*

The framework of the proposed method is shown in Fig.1. This framework contains two main parts: 1- prior information embedding and input data adjustment, and 2- IRC segmentation model. In the first part, by applying zero-padding, we provided the possibility of feeding OCT images to the neural network with their effective size. In addition, we employed expert knowledge to compensate for the side effects of zero-padding and also drive the network to learn the location of abnormalities more efficiently. In the second part, an extended structure of the U-Net model was implemented by applying a connection module between encoder and decoder parts, indicated with black dash dots in Fig.1, comprise of attention-gates (Schlemper et al., 2019) in the skip connection layers and atrous spatial pyramid pooling (ASPP) structure (L. C. Chen et al., 2018) in the bottleneck layer. In this architecture, through the attention gates, the network can focus more on the area where the cysts occur, and through the ASPP structure, by multi-scale field of view enlargement, it can obtain more efficient multi-scale contextual features without more decreasing the resolution of the feature map. In the following, each part will be explained in detail.

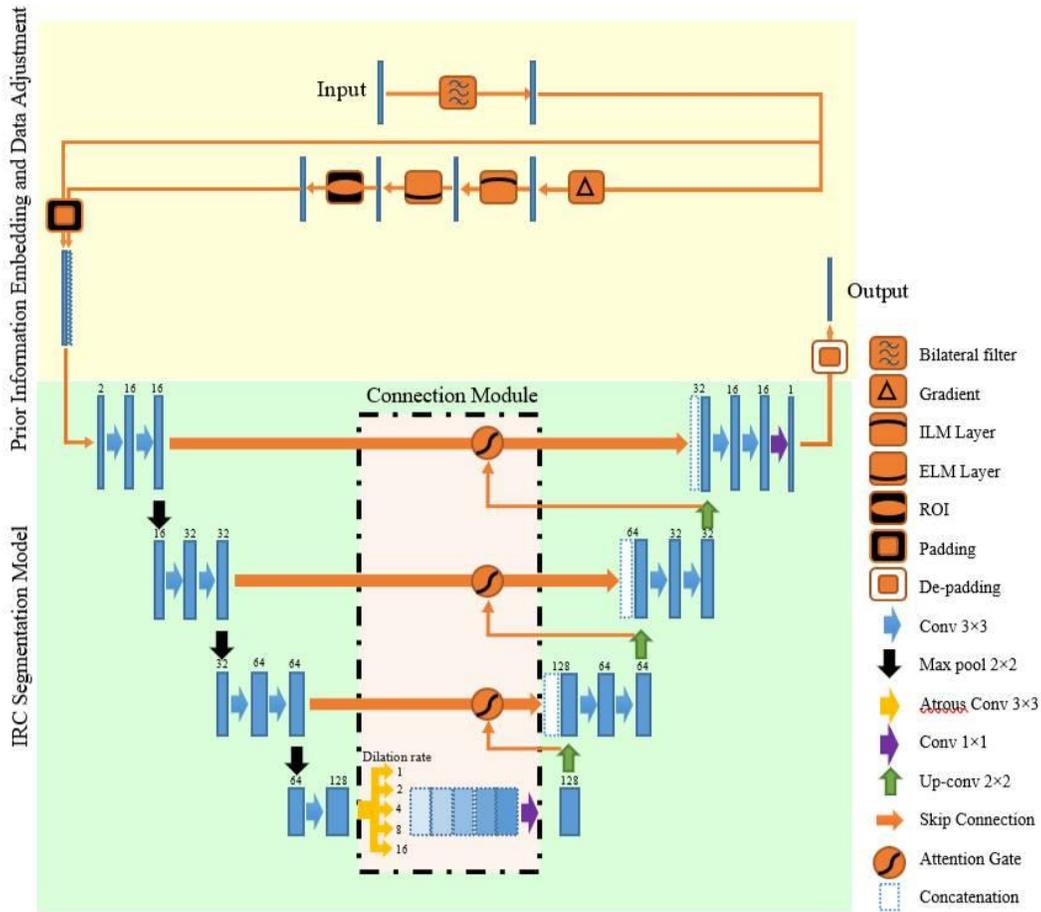

Fig.1.The block diagram of the proposed method.

## 2.1. Prior information embedding and input data adjustment

In this section, the first purpose is to prepare data set such that OCT images are feed into the neural network with an effective size and with keeping their aspect ratio instead of asymmetrically and blindly resizing to the equal dimensions; and the second purpose is to embed prior information as an extra channel to each input image. To achieve these goals, after enhancing the quality of the input images, first by applying an unsupervised segmentation method, an effective region of each image that is prone to the cyst occurrence named region of interest (ROI) image is extracted, then by determining the largest possible dimension in OCT images provided by different vendors as a reference dimension, both preprocessed and ROI images are dilated to this dimension by zero-padding. In the two following subsections, the specifics of this step are described in more detail.

*2.1.1. Denoising*

The varying degree of speckle-noise degrades the quality of OCT images (Yan et al., 2020). Since noisy images can reduce the segmentation performance, we employ conventional bilateral filtering to reduce speckle noise before segmentation; because this simple denoising model can smooth noisy scans and keep edge data with relatively low time complexity. By applying this filter to the input image $I(x)$, the denoised image $I^d(x)$ is computed as follows (Tomasi & Manduchi, 1998):

$$I^d(x) = k^{-1}(x) \sum_{x_i \epsilon \Omega} I(x_i) G_{\sigma_d}(\|x_i - x\|) G_{\sigma_r}(I(x_i) - I(x)) \qquad (1)$$

Where $\sigma_r$ and $\sigma_d$ are standard deviations of Gaussian functions named respectively as range and spatial parameters. $\|\cdot\|$ shows Euclidean distance; $\Omega$ indicates neighborhood pixels with center $x$, and $k = \sum_{x_i \epsilon \Omega} G_{\sigma_d}(\|x_i - x\|) G_{\sigma_r}(I(x_i) - I(x))$ is a normalization function.

As it is clear in equation (1), using bilateral filtering, each noisy image pixel $x$ enhanced by computing both Euclidean distance and intensity similarity of its nearby pixels and replacing its value with a weighted intensity average of close pixels. The result of applying this filter on a sample of OCT scan is shown in Fig.2 (a) and (b). As can be seen, this image filtering could enhance the quality of the input image properly (Ganjee et al., 2020).

*2.1.2. ROI segmentation and Zero padding*

One of the limitations of U-Net-based segmentation models is receiving the input data in the same size. This feature is challenging in cases such as vendor-independent IRC segmentation that images have different resolutions and dimensions. In this work, to preserve the originality of OCT images, zero-padding is used instead of the asymmetric input image resizing. In this way, the largest possible dimensions of an OCT image in horizontal and vertical coordinates are considered as reference dimensions, and then by applying zero-padding to each image, each OCT image is embedded in the reference dimension. In order to compensate for the side effects of zero-padding and also injecting the prior information of the IRC location to the FCN model, each input data is enriched by adding a second channel with the information of ROI in addition to the original image. This strategy not only suppresses irrelevant information in the input image but also improves another limitation of the U-Net model in learning where cysts can be located.

Based on expert knowledge, IRCs happen in a limited area from the internal limiting membrane (ILM) layer to the inner segment myeloid (ISM) layer of the retina. Since in SD-OCT images these two layers emerge with high contrast, a simple unsupervised graph-based method (Chiu et al., 2010) is applied for extraction of them. In this way, first, each OCT image is mapped to an undirected weighted graph and then desired layers are segmented by detecting two shortest graph paths using Dijkstra's algorithm. The weighted graph is constructed by mapping each image pixel to a graph node and defining each graph edge as a linkage between two neighbor pixels. The vertical gradient is used to obtain the weight of each edge because, in retinal OCT visualizing, pixel intensities from one layer to another layer vary in the vertical direction.

Hence, each edge weight $w_{ab}$ that connects two neighboring pixels $a$ and $b$ is obtained as follows (Chiu et al., 2010):

$$w_{ab} = 2 - (g_a + g_b) + w_{min} \qquad (2)$$

where $w_{min}$ is a constant that is set to a small value. $g_a$ and $g_b$ respectively indicate the vertical gradient values of the pixel a and b.

After constructing the weighted graph model, one of the desired layers is segmented by extracting the first shortest graph path using Dijkstra's algorithm (Dijkstra, 1959). Then, the constructed graph is cut into two subgraphs using this segmented layer. If the area above the cut consists of hyper-reflective data, the segmented layer is ISM, and hence the top subgraph is considered for ILM segmentation. Otherwise, it is ILM, and thus the bottom subgraph is investigated for ISM segmentation. by segmenting ILM and ISM layers, ROI is considered as the area restricted between these two layers. Fig.2 (c) and (d) show ROI segmentation in an enhanced OCT sample shown in Fig.2 (b) (Ganjee et al., 2020).

The ROI image is also padded with zero and then integrated with the corresponding zero-padded image to make a two-channel input image for the next processing.

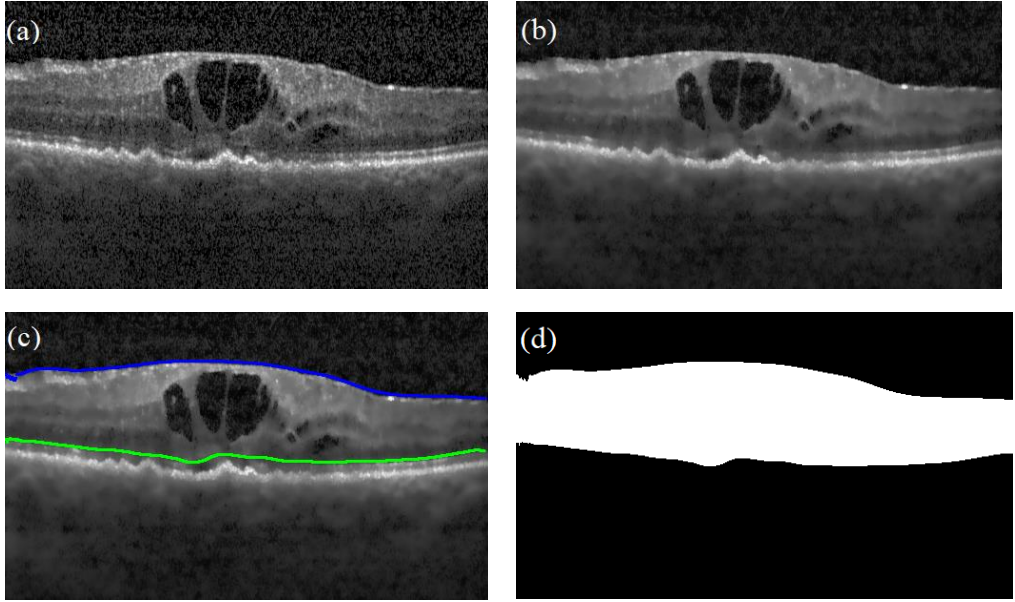

Fig.2.Preprocessing and ROI segmentation on a sample OCT image. (a) Original image, (b) Enhanced image, (c) ILM (blue) and ISM (green) segmentation, (d) ROI image.

## 2.2. The IRC segmentation model

U-Net is a well-known FCN model that is widely used for medical image segmentation (Ronneberger et al., 2015). Despite its success, due to the symmetric encoder-decoder architecture with skip connections that let to employ a combination of global and local features in image segmentation task, it has limitations in learning the location of the target object and increasing FOV without decreasing the resolution of the feature map. Hence, we proposed a connection module in the core of U-Net architecture, where encoder connects to decoder, that uses attention gates (Schlemper et al., 2019) in the skip connections and atrous spatial pyramid pooling (ASPP) (L. C. Chen et al., 2018) after the first convolution in bottleneck layer. By integrating attention gates in the skip connections, the network is driven to learn where lesions or target objects can happen and by embedding an atrous spatial pyramid pooling in the bottleneck layer, the receptive field at the high-level feature map is extended with various dilation rates that led to enriching the network with more multi-scale contextual features without more decreasing of feature map resolution. As is shown in Fig.1, our FCN model is composed of three parts: encoder, decoder, and connection module. The encoder part is a contracting path to capture contextual features at different resolutions that composed of multiple downsampling layers. Every downsampling layer consists of two convolutional layers followed by a max-pooling operation. The decoder part is an expanding path to localize target objects, including boundaries and contours. This part is composed of multiple up-sampling layers so that every up-sampling layer consists of two convolutional layers followed by a de-convolutional layer. And the connection module with attention gates and ASPP to concatenate symmetrical contextual and positional features obtained from convolution and de-convolution layers of encoder and decoder parts, respectively. Every attention gate in the connection module receives two inputs provided by the last convolution in each encoder layer ($X^l$) and the de-convolution in each decoder layer ($g$), and then computes the output $\hat{X}^l$ based on the attention coefficient $\alpha^l$ as follows:

$$\hat{X}^l = \alpha^l X^l, \qquad (3)$$

$$\alpha^l = \sigma_2\left(\psi^T\left(\sigma_1(W_x^T x^l + W_g^T g + b_{xg})\right) + b_\psi\right), \qquad (4)$$

where $\sigma_1$ and $\sigma_2$ are respectively sigmoid activation function and rectified linear unit (Relu). $\psi$, $W_x$ and $W_g$ are linear transformation that computed using $1 \times 1$ convolutions, and $b_{xg}$ and $b_\psi$ are bias terms. In fact, $\psi$, $W_x$, $W_g$, $b_{xg}$ and $b_\psi$ are attention gate parameters that are computed during the training process. These attention gates help the network to pass salient features through the skip connections before performing concatenation operations that lead to merge relevant features for further processing. Another structure used in the connection module is ASPP that employed multiple parallel atrous convolutions with different dilation rates. Standard convolution is a special mode of atrous convulsion. For the input feature map $x$ and convolution filter $w$, the standard convolution in each location $i$ of output feature map $y$ is computed as follows:

$$y[i] = \sum_k x[i+k]w[k], \qquad (5)$$

And the atrous convolution is defined as:

$$y[i] = \sum_k x[i+rk]w[k], \qquad (6)$$

where $r$ is a dilation rate that determines the number of zero values that should be added between two neighboring values of convolution kernel, standard convolution occurs when $r = 1$. Performing the atrous convolution instead of the standard convolution allows us to enlarge the field-of-view and capture more efficient multi-scale contextual feature maps. These atrous convolution results are concatenated to each other, and then by passing through a standard 1 * 1 convolution, multi-scale contextual features are integrated and send to the decoder part.

The proposed FCN model was trained end-to-end with the binary cross-entropy loss function. The specifics of training will describe in more detail in the experimental setup and measures subsection.

## 3. Experimental results

### 3.1. data set

Three publicly available databases namely OPTIMA (Wu et al., 2016), KERMANY (Kermany et al., 2018), and UMN (Rashno et al., 2018) were employed in this work. The OPTIMA dataset contains 30 OCT volumes divided into training and testing sets, each set containing 15 volumes that have been captured from CME subjects by employing four vendors including Cirrus, Topcon, Spectralis, and Nidek. Both training and testing sets are included with 3 OCT volumes from Nidek and 4 volumes from each of the three other vendors. The total number of scans provided in training and testing sets of this dataset is equal to 1560 and 909, respectively. In the UMN and KERMANY datasets, only Spectralis device was used for SD-OCT image capturing, and they respectively contain 725 and 256 scans that were imaged from DME patients. Two trained experts manually delineated the ground truths in the OPTIMA and UMN datasets. In contrast, three raters were employed to perform this task in the KERMANY dataset,. (Lu et al., 2019). To have a fair comprising with state-of-the-art methods, the UMN dataset and the training set of the OPTIMA dataset was used to train the network, and the KERMANY dataset and the testing set of the OPTIMA dataset were used for evaluations.

### 3.2. Experimental setup and Measures

In the first part of the proposed method, to denoise the input images, $\sigma_r$ is estimated adaptively by automatic extraction of a patch from the region above the ILM layer. To control the degree of image smoothing and edges blurring, the $\sigma_d$ parameter, which specifies the filter size, was set near the default value equal to 2 (Tomasi & Manduchi, 1998); to ROI segmentation, a weighted undirected graph was used to segment ILM and ISM layers. the weights of the graph were obtained using equation (2) with the minimum weight $w_{min}$ set to $10^{-5}$ (Chiu et al., 2010); And to unify the size of the images obtained from different vendors, all preprocessed images and their corresponding ROI images are passed through a padding module to construct two-channel images with the equal size of $640 * 1024 * 2$. Preprocessed and ROI images are included in the center of this reference dimension.

In the second part of the proposed method, the proposed extended U-Net model has two main components, the base U-Net model and the connection module. Although the higher number of convolutional layers in the base U-Net resulting in more accurate segmentation maps, it causes to increase in model parameters that can lead to overfitting and more computational cost. Hence, to have a tradeoff between the efficiency of the segmentation and computational cost, we chose four convolutional layers in the base U-Net model that extended feature dimensions from 16 in the first layer to 128 feature maps in the bottleneck; Consequently, the connection module has four layers: three layers with the attention gates and one layer with the ASPP. All of the convolution used in the base U-Net are $3 \times 3$, and in the gate layer of the connection module, three $1 \times 1$ convolutions have been employed; in the ASPP layer of the connection module, five atrous convolutions with the size of $3 \times 3$ and dilation rates {1, 2, 4, 8, 16}, and one $1 \times 1$ standard convolution were considered. To avoid overfitting, dropout rates were set to 0.1 in the first and second layers, and 0.2 in the third and fourth layers. And finally, our extended U-Net model was trained with a batch size of 10 and iterated for 100 epochs from scratch using binary cross-entropy loss function. The weights were updated using the Adam optimizer with default parameters (Kamnitsas et al., 2015), and the learning rate was empirically set to $10^{-3}$.

The output of the proposed method is a probability map that indicates cyst prediction scores for each pixel of the input image. In the test phase, this probability map is thresholded with the threshold value 0.5 to obtain the binary mask of the segmented IRC regions.

Three well-known measures namely recall, precision, and dice coefficient (DC) were used for performance evaluation of the proposed methods. These measures are defined as follows:

$$recall = \frac{TP}{TP+FN}, precision = \frac{TP}{TP+FP}, \qquad (7)$$

$$DC = 2 \cdot \frac{|SR \cap GT|}{|SR|+|GT|}, \qquad (8)$$

where $TP$ and $FP$ are respectively the number of image pixels that are correctly and wrongly classified as cystic pixels; and $FN$ is cystic pixels that are not extracted by the automatic method.

In equation (8), $DC$ is computed between the segmentation results provided by the automatic method ($SR$) and the ground truths ($GT$) for each OCT image in test set.

### 3.3. Results

The efficiency of the presented method was evaluated quantitatively and qualitatively across two public OPTIMA and KERMANY datasets. Table 1 shows the inter-observer variability (IOV) across four vendors in the OPTIMA dataset. Evaluation of this measure in the KERMANY dataset is not possible because three ground truths of this dataset are not available separately and only the intersection of them have been published (Lu et al., 2019). As can be seen, manually grading IRC regions is a difficult task even for human experts because of the low contrast and inhomogeneous changes of intensity in these images.

**Table 1. inter-observer variability (IOV) in terms of mean (standard deviation) DC values in four vendors of the OPTIMA dataset.**

|     | Cirrus | Topcon | Spectralis | Nidek |
|-----|--------|--------|------------|-------|
| IOV | 0.95 (0.01) | 0.91 (0.02) | 0.87 (0.12) | 0.85 (0.08) |

Table 2 shows the results of the presented method on the test set of the OPTIMA using mean (standard deviation) recall, precision, and dice coefficient measures across manual segmentations prepared by expert 1 ($GT1$), expert 2 ($GT2$), and the intersection of them ($GT1 \cap GT2$). As it is reported in this Table, the highest performance of the proposed method was obtained on SD-OCT images from Spectralis device with mean recall, precision, and dice coefficient rates of 0.87, 0.95, and 0.87 respectively. And the lowest efficiency is related to the Nidek SD-OCT image set with the mean DC result of 0.69 due to the more intensity variations and ambiguous boundaries compared to the other OCT image sets.

**Table 2. Mean (standard deviation) recall, precision, and dice coefficient results of the presented method on OPTIMA test images separately for every four vendors.**

|            | $GT1$ | | | $GT2$ | | | $GT1 \cap GT2$ | | |
|------------|--------|-----------|----|--------|-----------|----|--------|-----------|----|
|            | Recall | Precision | DC | Recall | Precision | DC | Recall | Precision | DC |
| Cirrus     | 0.86 (0.03) | 0.84 (0.12) | 0.77 (0.11) | 0.86 (0.02) | 0.85 (0.12) | 0.77 (0.11) | 0.88 (0.02) | 0.84 (0.12) | 0.78 (0.11) |
| Topcon     | 0.85 (0.21) | 0.83 (0.28) | 0.76 (0.15) | 0.88 (0.21) | 0.81 (0.28) | 0.76 (0.16) | 0.90 (0.21) | 0.79 (0.28) | 0.76 (0.16) |
| Spectralis | 0.85 (0.19) | 0.96 (0.20) | 0.86 (0.18) | 0.86 (0.16) | 0.96 (0.21) | 0.87 (0.17) | 0.87 (0.14) | 0.95 (0.24) | 0.87 (0.18) |
| Nidek      | 0.88 (0.06) | 0.69 (0.16) | 0.73 (0.18) | 0.87 (0.05) | 0.67 (0.13) | 0.72 (0.16) | 0.92 (0.06) | 0.60 (0.09) | 0.69 (0.14) |
| **all**    | **0.86 (0.12)** | **0.83 (0.19)** | **0.78 (0.16)** | **0.87 (0.11)** | **0.82 (0.19)** | **0.78 (0.15)** | **0.84 (0.11)** | **0.79 (0.18)** | **0.78 (0.15)** |

Table 3 shows the performance of the proposed method in comparison with the participants in the OPTIMA cyst challenge (*Optima cyst segmentation challenge*, 2015), some other recent methods published in (Rashno et al., 2018), (Girish et al., 2019; Gopinath & Sivaswamy, 2019; Rashno et al., 2018; Venhuizen et al., 2018), and the most recent method presented by us (Ganjee et al., 2020) on OPTIMA image set, and Table 4 shows our results on KERMANY dataset compared to our previous method (Ganjee et al., 2020) and other recent method published in (Lu et al., 2019). To show the efficiency of our extended U-Net model, three U-Net-based architectures including standard U-Net, U-Net with attention gates (U-Net+AG), and U-Net with ASPP (U-Net+ ASPP) were implemented in the framework of the proposed method and their results were also included in Tables 3 and 4. As it is seen, on the OPTIMA dataset the proposed method successes to outperform the compared methods with a mean DC value of 0.78 on all ground truths. And on the KERMANY dataset, the proposed method can operate better than our previous method with 3% improvement of dice coefficient rate, although it is slightly inferior in comparison with the Lu et al.(Lu et al., 2019) method due to the large difference in the size of the training set and also vendor-specific training strategy in (Lu et al.,

2019). It should be mentioned, the papers published in (Esmaeili et al., 2016) and (Rashno et al., 2018) were only used the train set of the Spectralis device to evaluate their method on the OPTIMA dataset.

Another point that is understood from Table 3 is that the implementation of the proposed method with the standard U-Net can obtain results as well as the method proposed by Grish et al.(Girish et al., 2019), that employed standard U-Net with asymmetric resized input data. It shows that our strategy in injecting data with different sizes to the network while making the proposed method robust against input image size did not cause our method to lose efficiency.

Table 3. The presented method comparison in terms of mean (standard deviation) of DC on OPTIMA dataset with the optima cyst challenge participants (Optima cyst segmentation challenge, 2015), recent deep-based (Girish et al., 2019; Gopinath & Sivaswamy, 2019; Lu et al., 2019; Venhuizen et al., 2018), and unsupervised methods ( Ganjee et al., 2020; Rashno et al., 2018), and the presented method implementation with three U-Net based models.

| Method | Approach | year | GT1 | GT2 | GT1 ∩ GT2 |
|---|---|---|---|---|---|
| Oguz et al.(Oguz et al., 2016) | Unsupervised | 2016 | 0.48 (0.25) | 0.48 (0.22) | 0.48 (0.22) |
| Esmaeili et al. (Esmaeili et al., 2016) | | 2016 | 0.46 (0.25) | 0.45 (0.24) | 0.45 (0.25) |
| Rashno et al. (Rashno et al., 2018) | | 2018 | 0.70 (0.10) | 0.71 (0.11) | 0.72 (0.10) |
| Ganjee et al.(Ganjee et al., 2020) | | 2020 | 0.74 (0.14) | 0.73 (0.14) | 0.74 (0.14) |
| de Sisternes et al. (de Sisternes et al., 2015) | Supervised | 2015 | 0.64 (0.14) | 0.63 (0.14) | 0.65 (0.15) |
| Venhuizen et al.(Venhuizen, F., van Grinsven, M. J., van Ginneken, B., Hoyng, C. C., Theelen, T., Sanchez, 2016) | | 2016 | 0.56 (0.20) | 0.55 (0.22) | 0.54 (0.20) |
| Haritz et al.(Gopinath & Sivaswamy, 2016) | | 2016 | 0.14 (0.08) | 0.14 (0.08) | 0.14 (0.08) |
| Venhuizen et al.(Venhuizen et al., 2018) | Supervised (deep) | 2018 | - | - | 0.74 (0.16) |
| Grish et al.(Girish et al., 2019) | | 2018 | 0.71 (0.20) | 0.72 (0.19) | 0.72 (0.19) |
| Gopinath et al.(Gopinath & Sivaswamy, 2019) | | 2018 | 0.67 (0.17) | 0.68 (0.17) | 0.69 (0.18) |
| Presented method with U-Net | | 2021 | 0.71 (0.16) | 0.71 (0.16) | 0.72 (0.16) |
| Presented method with U-Net+AG | | 2021 | 0.74 (0.14) | 0.74 (0.14) | 0.74 (0.14) |

Table 4. The presented method comparison in terms of mean (standard deviation) of DC on KERMANY dataset with recent deep-based method (Lu et al., 2019) , our previous unsupervised method (Ganjee et al., 2020), and the presented method implementation with three U-Net based models.

| Method | Approach | year | GT |
|---|---|---|---|
| Ganjee et al.(Ganjee et al., 2020) | Unsupervised | 2020 | 0.79 (0.15) |
| Lu et al. (Lu et al., 2019) | Supervised (deep) | 2019 | 0.82 (0.15) |
| Presented method with U-Net | | 2021 | 0.78 (0.17) |
| Presented method with U-Net+AG | | 2021 | 0.80 (0.14) |
| Presented method with U-Net+ASPP | | 2021 | 0.79 (0.15) |
| Presented method with U-Net+connection module | | 2021 | **0.81(0.15)** |

Qualitative evaluation of the proposed method is presented in Fig. 3-7 by employing four samples provided by each vendor in the OPTIMA dataset (Fig.3-6) and a sample from the KERMANY dataset (Fig.7). In each Fig.3-6, (a) shows the preprocessed image; (b) and (c) show the first and the second ground truth provided by the expert1 and expert2, in magenta and cyan colors, respectively; and in the third to sixth rows of Fig.3-6, each pair (d)-(e), (f)-(g), (h)-(i), and (j)-(k) show respectively the probability map and the final segmentation map obtained from the implementation of the proposed method with standard U-Net (red color), U-Net+AG (blue color), U-Net+ASPP (yellow color), and U-Net+connection module (green color). In Fig.7, for the KERMANY sample, (b) show the intersection of ground truths in white color because separate grader maps are not published for this dataset; and the second to fourth rows show the result of the proposed method with three mentioned U-Net based models and the proposed method with connection module. In qualitative evaluation, samples were chosen that can indicate the cysts segmentation challenges including segmentation of OCT images with variable cyst size, OCT images with low contrast cysts, and OCT images with cysts located near each other with indistinct boundary.

OCT samples containing various sizes of lesions are seen in all figures except Fig. 6, as it is observed the proposed model+connection module is more efficient in segmenting cysts with different sizes compared to the other models, especially in Fig. 5 and 7 where small cysts appear with low contrast. Samples with low contrast cysts are also seen in Fig. 6. As it is obvious, due to employing more effective contextual features through enlargement of the field of view, the presented method and U-Net+ASPP are more successful compared to the standard U-Net and U-Net+AG models, in addition, in comparison with the U-Net+ASPP, the segmentation result provided by the proposed method with connection module has a closer correlation with both ground truths. Fig.3 and 4 contain samples with ill-defined cysts boundary located near each other. As it is seen in these figures, totally the presented method has better performance in segmenting IRC regions up to their true boundary with the least false positive detection. Although in Fig.4, it seems that U-Net+AG has more separability power in detecting the boundary between cyst areas, this model could not delineate lesions up to their true boundary. In addition, as it is clear, there is a big disagreement between two graders in this sample so that the cyst map provided by the proposed method is more correlated with the second ground truth (shown with cyan color in Fig.4(c)) while U-Net+AG result has more correlation with the first ground truth (shown with magenta color in Fig.4(b)).

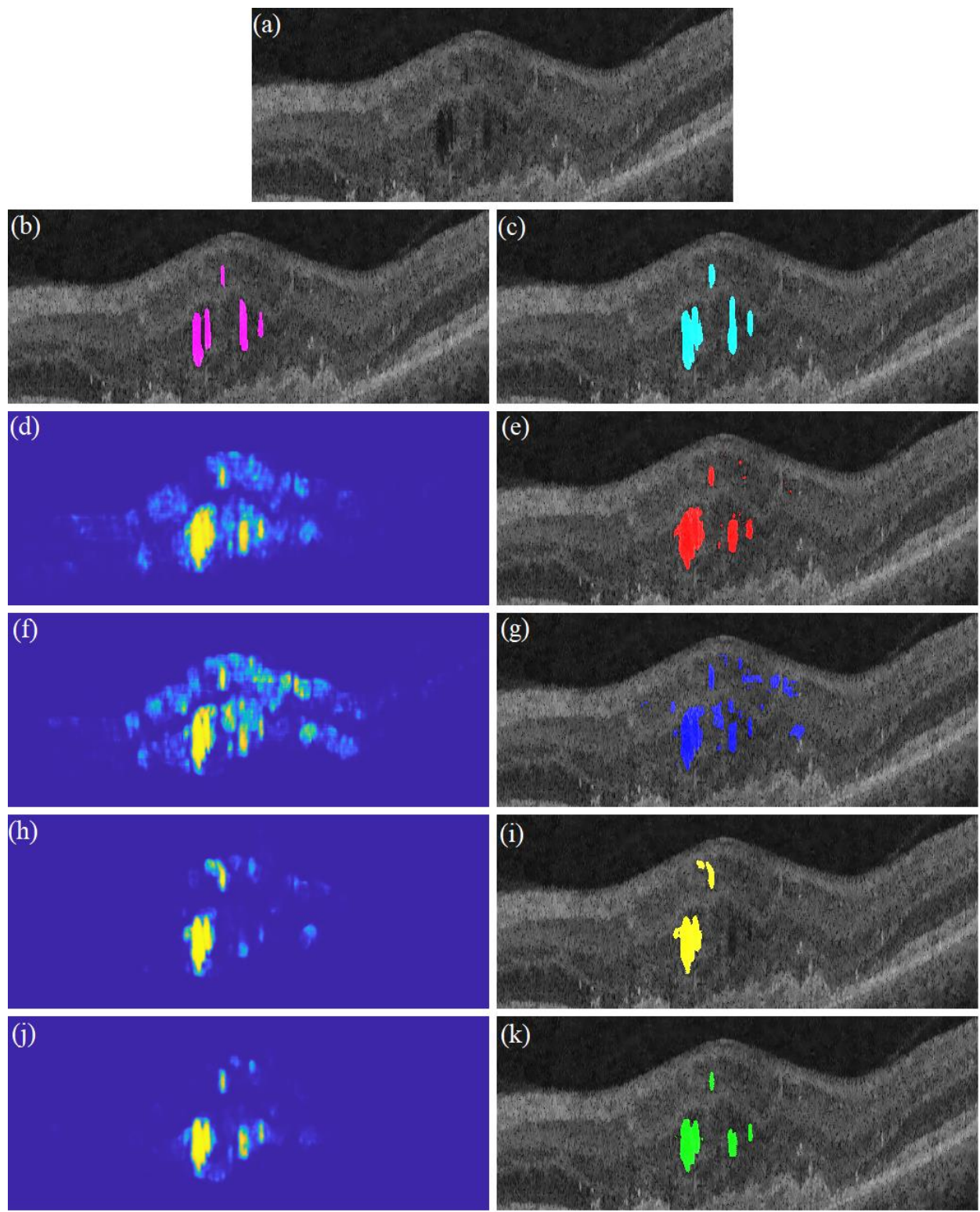

Fig. 3. Cirrus sample qualitative results using various implementations of the presented method with, U-Net, U-Net+AG, U-Net+ASPP, and U-Net+connection module. (a) preprocessed Cirrus sample, (b) GT1, (c) GT2, (d) U-Net probability map output, (e) U-Net segmentation result, (f) U-Net+AG probability map output, (g) U-Net+AG segmentation result, (h) U-Net+ASPP probability map output, (i) U-Net+ASPP segmentation result, (j)The probability map output of the proposed method+connection module, and (k) The proposed method+connection module segmentation result.

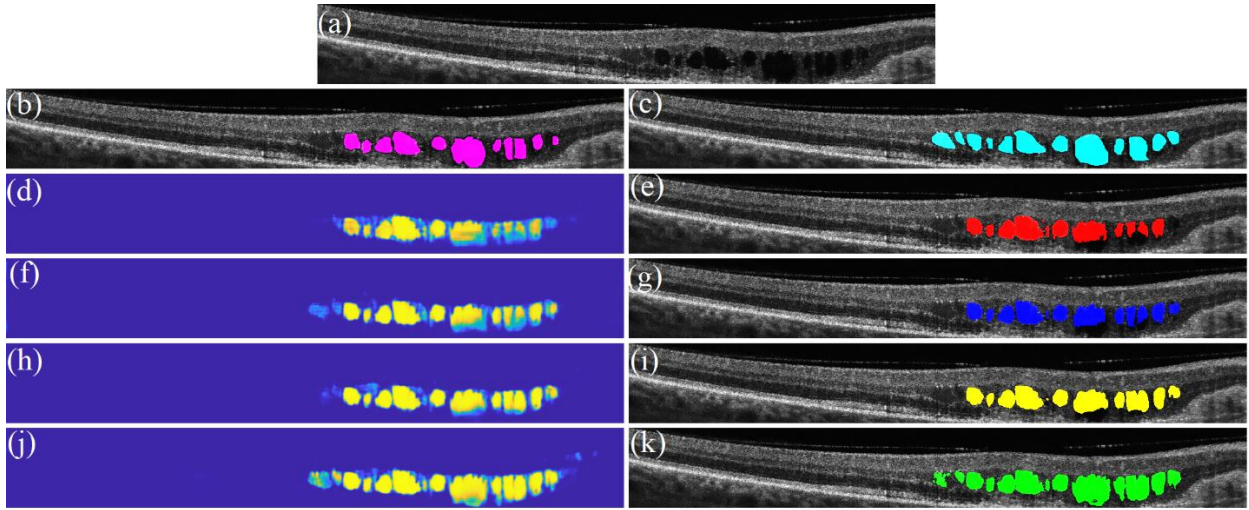

Fig. 4. Nidek sample qualitative results using various implementations of the presented method with, U-Net, U-Net+AG, U-Net+ASPP, and U-Net+connection module. (a) preprocessed Nidek sample, (b) GT1, (c) GT2, (d) U-Net probability map output, (e) U-Net segmentation result, (f) U-Net+AG probability map output, (g) U-Net+AG segmentation result, (h) U-Net+ASPP probability map output, (i) U-Net+ASPP segmentation result, (j)The probability map output of the proposed method+connection module, and (k) The proposed method+connection module segmentation result.

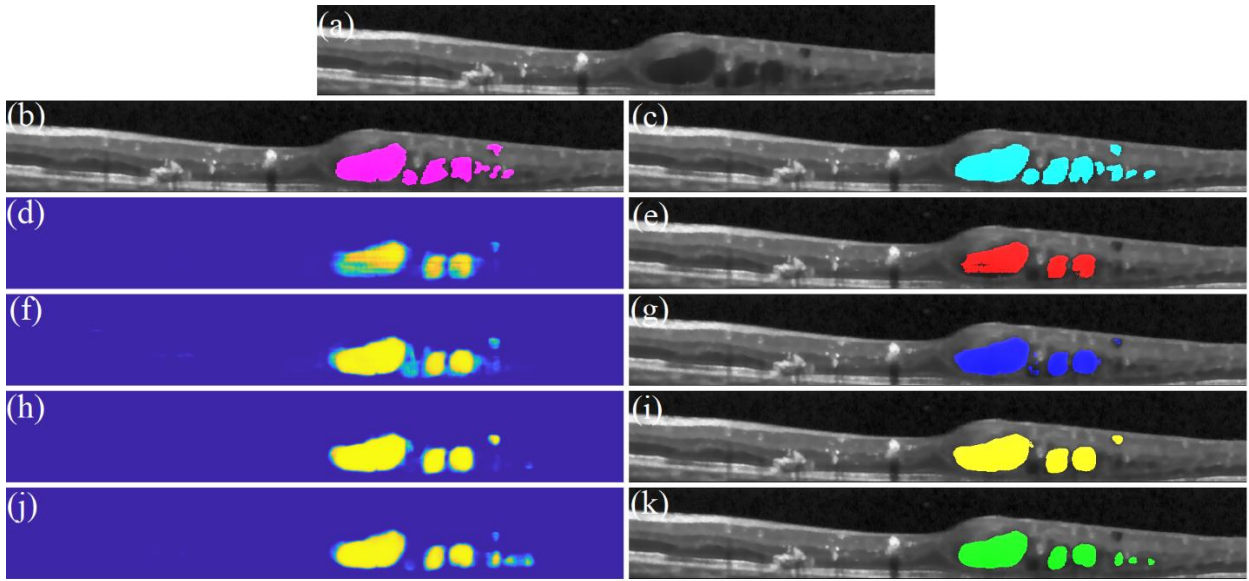

Fig. 5. Spectralis sample qualitative results using various implementations of the presented method with, U-Net, U-Net+AG, U-Net+ASPP, and U-Net+connection module. (a) preprocessed Spectralis sample, (b) GT1, (c) GT2, (d) U-Net probability map output, (e) U-Net segmentation result, (f) U-Net+AG probability map output, (g) U-Net+AG segmentation result, (h) U-Net+ASPP probability map output, (i) U-Net+ASPP segmentation result, (j)The probability map output of the proposed method+connection module, and (k) The proposed method+connection module segmentation result.

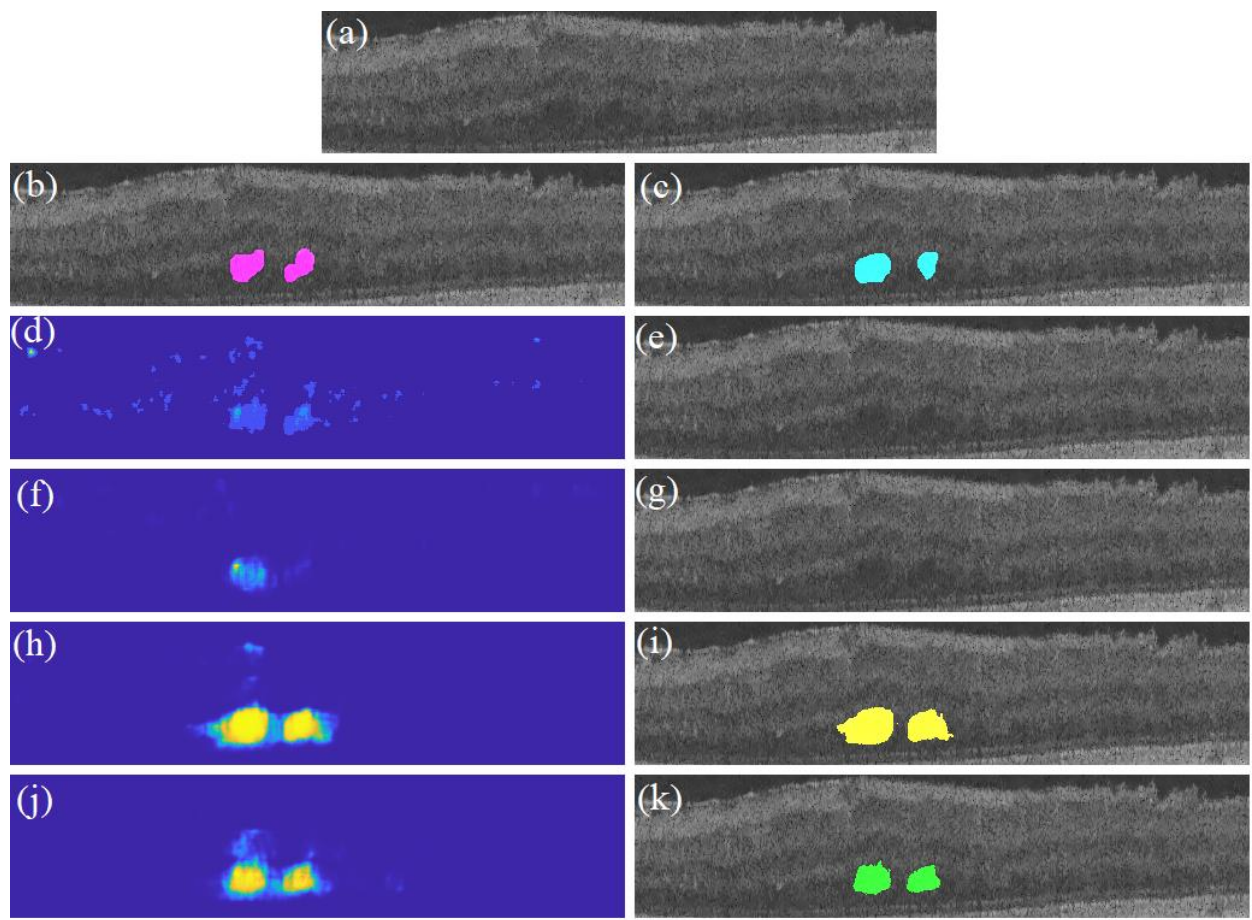

Fig. 6. Topcon sample qualitative results using various implementations of the presented method with, U-Net, U-Net+AG, U-Net+ASPP, and U-Net+connection module. (a) preprocessed Topcon sample, (b) GT1, (c) GT2, (d) U-Net probability map output, (e) U-Net segmentation result, (f) U-Net+AG probability map output, (g) U-Net+AG segmentation result, (h) U-Net+ASPP probability map output, (i) U-Net+ASPP segmentation result, (j)The probability map output of the proposed method+connection module, and (k) The proposed method+connection module segmentation result.

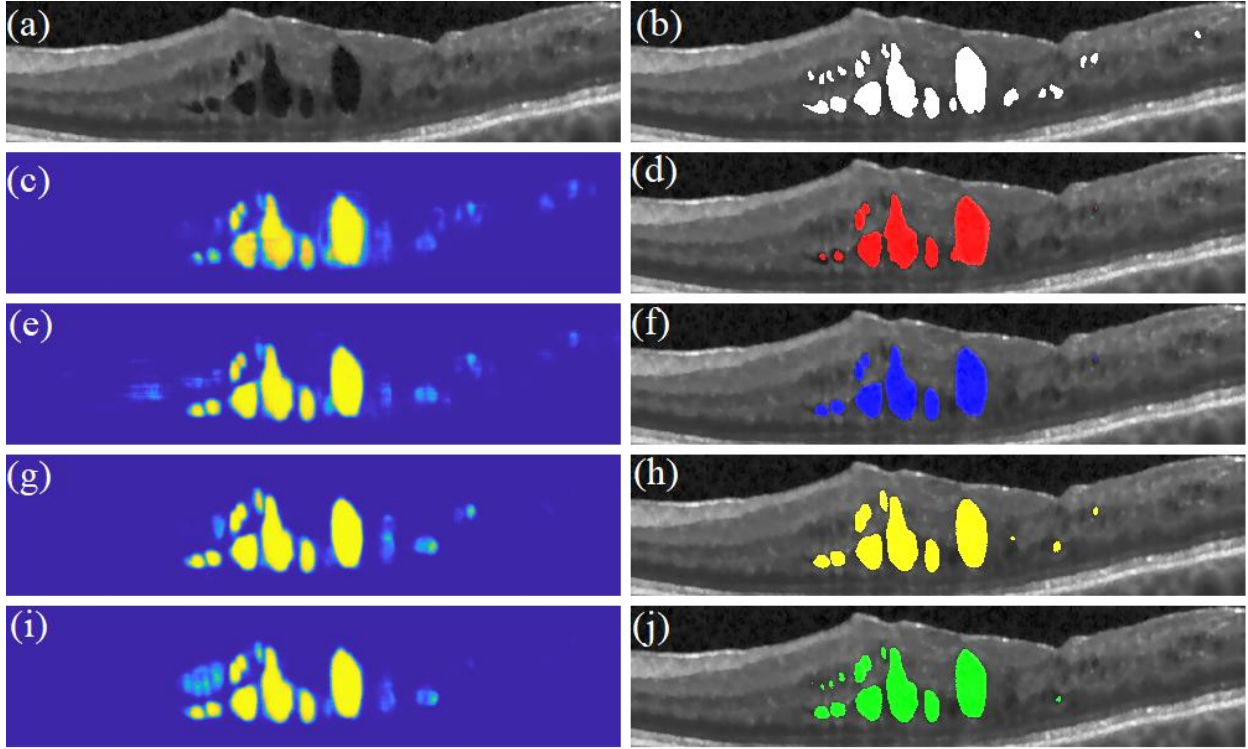

Fig. 7. KERMANY sample qualitative results using various implementations of the presented method with, U-Net, U-Net+AG, U-Net+ASPP, and U-Net+connection module. (a) preprocessed KERMANY sample, (b) GT1 ∩ GT2 ∩ GT3, (c) U-Net probability map output, (d) U-Net segmentation result, (e) U-Net+AG probability map output, (f) U-Net+AG segmentation result, (g) U-Net+ASPP probability map output, (h) U-Net+ASPP segmentation result, (i)The probability map output of the proposed method+connection module, and (j) The proposed method+connection module segmentation result.

*4. Discussion*

U-Net is a well-known model that is widely used in medical image segmentation due to its symmetric structure compose of an encoder-decoder with skip connections that seen global information (contextual features) along with local information (localization). However, U-Net based models have limitations in learning effective information such as shape, size, and location (i.e. where target object can happen), information that exists as prior knowledge in the context of medical image processing problems, and if it properly injected into the U-Net model, more accurate and better results will be provided (Jurdi et al., 2020).

In this paper, we tried to use this kind of information in our proposed U-Net based approach so that the network can take into account these effective features during the learning process compared to the standard U-Net model. In this way, we proposed a two-step approach that can receive this useful information both in the data and model levels. At the data level, we extended the number of input channels so that the first channel corresponds to the original image and the extra channels can incorporate domain expertise with the data. By this strategy, we also could manage the limitation of the U-Net model in receiving data of the same size; one of the challenges in deep-based vendor-independent methods. And at the model level, we extended the standard U-Net model with a connection module that enriched with attention gates and ASPP that drive the network to find salient regions, avoid irrelevant features and learn where a target object can happen, and also see more meaningful contextual features to segment lesions with various size and more accurate boundary.

As the results showed, our proposed method is generalizable across different sites and vendors; because the presented method could obtain accurate results on the unseen dataset ( KERMANY dataset) provided from a different site, hence our method is generalizable across different sites; and it also can operate successfully on images provided by different vendors, thus it is generalizable across different vendors. In comparison with the previous vendor-independent deep learning-based methods, vendor independency has been evaluated more precisely in our proposed method because we did not asymmetrically resize input images, and therefore we did not lose small cysts or we did not connect nearby cysts. And the ability of the segmentation model was evaluated by the original ground truth, not the resized version. In addition, any post-processing step was not employed in our proposed method that confirms the effectiveness of the prior information embedding as an extra input channel and connection module as a way to learn more effective features in managing false positive detections.

## 5. Conclusion and future works

Here, we presented a vendor-independent method based on the U-Net architecture that can operate on SD-OCT images obtained from various devices in their original size and provide acceptable results. Our solution summarizes in two parts: 1- prior information embedding and input data adjustment, 2- IRC segmentation model. In the first part, the asymmetrically resizing problem was solved by zero-padding; and then by adding an extra input channel with the information of target object location, both the model focus is increased, and the impact of outliers on the model learning process is decreased. And in the second part, we applied a connection module comprise of attention gates in the skip connection layers and atrous spatial pyramid pooling in the bottleneck layer to the core of the U-Net to transfer more useful and meaningful information from the encoder to the decoder part.

Two public datasets named OPTIMA with images provided by different OCT devices and KERMANY with Spectralis images from DME subjects were used to evaluate the proposed method. Using the ground truths intersection, our method could outperform all of the previous vendor-independent IRC segmentation methods by a mean Dice value of 0.78 and 0.81 on OPTIMA, and KERMANY datasets, respectively.

It is expected that injecting more powerful channels included with the domain expertise such as weighted ROI mask or a map with more effective information about the correlation between lesions and retinal layer structures to the network, can provide a more helpful introduction of the cystoid features to the network. It will be investigated in future work.

## References

Bogunovic, H., Venhuizen, F., Klimscha, S., Apostolopoulos, S., Bab-Hadiashar, A., Bagci, U., Beg, M. F., Bekalo, L., Chen, Q., Ciller, C., Gopinath, K., Gostar, A. K., Jeon, K., Ji, Z., Kang, S. H., Koozekanani, D. D., Lu, D., Morley, D., Parhi, K. K., … Schmidt-Erfurth, U. (2019). RETOUCH: The Retinal OCT Fluid Detection and Segmentation Benchmark and Challenge. *IEEE Transactions on Medical Imaging*, *38*(8), 1858–1874. https://doi.org/10.1109/TMI.2019.2901398

Chen, L. C., Zhu, Y., Papandreou, G., Schroff, F., & Adam, H. (2018). Encoder-decoder with atrous separable convolution for semantic image segmentation. *ArXiv*, 801–818.

Chen, X., Niemeijer, M., Zhang, L., Lee, K., Abramoff, M. D., & Sonka, M. (2012). Three-dimensional segmentation of fluid-associated abnormalities in retinal OCT: Probability constrained graph-search-graph-cut. *IEEE Transactions on Medical Imaging*, *31*(8), 1521–1531. https://doi.org/10.1109/TMI.2012.2191302

Chiu, S. J., Allingham, M. J., Mettu, P. S., Cousins, S. W., Izatt, J. A., & Farsiu, S. (2015). Kernel regression based segmentation of optical coherence tomography images with diabetic macular edema. *Biomedical Optics Express*, *6*(4), 1172–1194. https://doi.org/10.1364/boe.6.001172

Chiu, S. J., Li, X. T., Nicholas, P., Toth, C. A., Izatt, J. A., & Farsiu, S. (2010). Automatic segmentation of seven retinal layers in SDOCT images congruent with expert manual segmentation. *Optics Express*, *18*(18), 19413–19428. https://doi.org/10.1364/oe.18.019413

de Sisternes, L., Hong, J., Leng, T., & Rubin, D. L. (2015). A machine learning approach for device-independent automated segmentation of retinal cysts in spectral domain optical coherence tomography images. *Proceeding Optima Challenge-MICCAI*.

Dijkstra, E. W. (1959). A note on two problems in connexion with graphs. *Numerische Mathematik*, *1*(1), 269–271. https://doi.org/10.1007/BF01386390

Esmaeili, M., Dehnavi, A., Rabbani, H., & Hajizadeh, F. (2016). Three-dimensional segmentation of retinal cysts from spectral-domain optical coherence tomography images by the use of three-dimensional curvelet based K-SVD. *Journal of Medical Signals and Sensors*, *6*(3), 166–171.

Ganjee, R., Ebrahimi Moghaddam, M., & Nourinia, R. (2020). An unsupervised hierarchical approach for automatic intra-retinal cyst segmentation in spectral-domain optical coherence tomography images. *Medical Physics*, *47*(10), 4872–4884. https://doi.org/10.1002/mp.14361

Ganjee, R., Moghaddam, M. E., & Nourinia, R. (2018). Automatic segmentation of abnormal capillary nonperfusion regions in optical coherence tomography angiography images using marker-controlled watershed algorithm. *Journal of Biomedical Optics*, *23*(09), 096006. https://doi.org/10.1117/1.jbo.23.9.096006

Girish, G. N., R. Kothari, A., & Rajan, J. (2018). Marker controlled watershed transform for intra-retinal cysts segmentation from optical coherence tomography B-scans. *Pattern Recognition Letters*. https://doi.org/10.1016/j.patrec.2017.12.019

Girish, G. N., Thakur, B., Chowdhury, S. R., Kothari, A. R., & Rajan, J. (2019). Segmentation of intra-retinal cysts

from optical coherence tomography images using a fully convolutional neural network model. *IEEE Journal of Biomedical and Health Informatics*, *23*(1), 296–304. https://doi.org/10.1109/JBHI.2018.2810379

González, A., Remeseiro, B., Ortega, M., Penedo, M. G., & Charlón, P. (2013). Automatic cyst detection in OCT retinal images combining region flooding and texture analysis. *Proceedings - IEEE Symposium on Computer-Based Medical Systems*, 397–400. https://doi.org/10.1109/CBMS.2013.6627825

Gopinath, K., & Sivaswamy, J. (2016). Domain knowledge assisted cyst segmentation in OCT retinal images. *ArXiv Preprint ArXiv:1612.02675*.

Gopinath, K., & Sivaswamy, J. (2019). Segmentation of Retinal Cysts from Optical Coherence Tomography Volumes Via Selective Enhancement. *IEEE Journal of Biomedical and Health Informatics*, *23*(1), 273–282. https://doi.org/10.1109/JBHI.2018.2793534

Jurdi, R. El, Petitjean, C., Honeine, P., & Abdallah, F. (2020). Bb-unet: U-net with bounding box prior. *IEEE Journal on Selected Topics in Signal Processing*, *14*(6), 1189–1198. https://doi.org/10.1109/JSTSP.2020.3001502

Kamnitsas, K., Chen, L.-C., Ledig, C., Rueckert, D., & Glocker, B. (2015). Multi-Scale 3D Convolutional Neural Networks for Lesion Segmentation in Brain MRI. *Proceedings of MICCAI-ISLES 2015*, *13*, 13–16. www.isles-challenge.org

Kermany, D., Zhang, K., & Goldbaum, M. (2018). Labeled optical coherence tomography (oct) and chest X-ray images for classification. *Mendeley Data*, *2*. https://data.mendeley.com/datasets/rscbjbr9sj/2#file-9e8f7acf-7d3a-487f-8eb5-0bd3255b9685

Lee, C. S., Tyring, A. J., Deruyter, N. P., Wu, Y., Rokem, A., & Lee, A. Y. (2017). Deep-learning based, automated segmentation of macular edema in optical coherence tomography. *Biomedical Optics Express*, *8*(7), 3440–3448. https://doi.org/10.1364/boe.8.003440

Lu, D., Heisler, M., Lee, S., Ding, G. W., Navajas, E., Sarunic, M. V., & Beg, M. F. (2019). Deep-learning based multiclass retinal fluid segmentation and detection in optical coherence tomography images using a fully convolutional neural network. *Medical Image Analysis*, *54*, 100–110. https://doi.org/10.1016/j.media.2019.02.011

Moschos, T. G. R. and M. M. (2008). Cystoid macular edema. *Clinical Ophthalmology (Auckland, NZ)*, *2*(4), 919–930.

Nazir, T., Irtaza, A., Shabbir, Z., Javed, A., Akram, U., & Mahmood, M. T. (2019). Diabetic retinopathy detection through novel tetragonal local octa patterns and extreme learning machines. *Artificial Intelligence in Medicine*, *99*, 101695. https://doi.org/10.1016/j.artmed.2019.07.003

Oguz, I., Zhang, L., Abràmoff, M. D., & Sonka, M. (2016). Optimal retinal cyst segmentation from OCT images. *Medical Imaging 2016: Image Processing*, *9784*, 97841E-97841E. https://doi.org/10.1117/12.2217355

*Optima cyst segmentation challenge*. (2015). https://optima.meduniwien.ac.at/research/challenges/

Quellec, G., Lee, K., Dolejsi, M., Garvin, M. K., Abràmoff, M. D., & Sonka, M. (2010). Three-dimensional analysis of retinal layer texture: Identification of fluid-filled regions in SD-OCT of the macula. *IEEE Transactions on Medical Imaging*, *29*(6), 1321–1330. https://doi.org/10.1109/TMI.2010.2047023

Rashno, A., Koozekanani, D. D., Drayna, P. M., Nazari, B., Sadri, S., Rabbani, H., & Parhi, K. K. (2018). Fully Automated Segmentation of Fluid/Cyst Regions in Optical Coherence Tomography Images with Diabetic Macular Edema Using Neutrosophic Sets and Graph Algorithms. *IEEE Transactions on Biomedical Engineering*, *65*(5), 989–1001. https://doi.org/10.1109/TBME.2017.2734058

Ronneberger, O., Fischer, P., & Brox, T. (2015). U-net: Convolutional networks for biomedical image segmentation. *Lecture Notes in Computer Science (Including Subseries Lecture Notes in Artificial Intelligence and Lecture Notes in Bioinformatics)*, *9351*, 234–241. https://doi.org/10.1007/978-3-319-24574-4_28

Roy, A. G., Conjeti, S., Karri, S. P. K., Sheet, D., Katouzian, A., Wachinger, C., & Navab, N. (2017). ReLayNet: retinal layer and fluid segmentation of macular optical coherence tomography using fully convolutional networks. *Biomedical Optics Express*, *8*(8), 3627–3642. https://doi.org/10.1364/boe.8.003627

Roychowdhury, S., Koozekanani, D. D., Radwan, S., & Parhi, K. K. (2013). Automated localization of cysts in diabetic macular edema using optical coherence tomography images. *Proceedings of the Annual International Conference of the IEEE Engineering in Medicine and Biology Society, EMBS*, 1426–1429. https://doi.org/10.1109/EMBC.2013.6609778

Schlegl, T., Waldstein, S. M., Bogunovic, H., Endstraßer, F., Sadeghipour, A., Philip, A. M., Podkowinski, D., Gerendas, B. S., Langs, G., & Schmidt-Erfurth, U. (2018). Fully Automated Detection and Quantification of Macular Fluid in OCT Using Deep Learning. *Ophthalmology*, *125*(4), 549–558. https://doi.org/10.1016/j.ophtha.2017.10.031

Schlegl, T., Waldstein, S. M., Vogl, W. D., Schmidt-Erfurth, U., & Langs, G. (2015). Predicting semantic descriptions from medical images with convolutional neural networks. *Lecture Notes in Computer Science (Including Subseries Lecture Notes in Artificial Intelligence and Lecture Notes in Bioinformatics)*, *9123*, 437–448. https://doi.org/10.1007/978-3-319-19992-4_34

Schlemper, J., Oktay, O., Schaap, M., Heinrich, M., Kainz, B., Glocker, B., & Rueckert, D. (2019). Attention gated networks: Learning to leverage salient regions in medical images. *Medical Image Analysis*, *53*, 197–207. https://doi.org/10.1016/j.media.2019.01.012.

Tomasi, C., & Manduchi, R. (1998). Bilateral filtering for gray and color images. *Proceedings of the IEEE International Conference on Computer Vision*, 839–846.

Venhuizen, F., van Grinsven, M. J., van Ginneken, B., Hoyng, C. C., Theelen, T., Sanchez, C. I. (2016). Fully automated segmentation of intraretinal cysts in 3D optical coherence tomography. *Invest. Ophthalmol. Vis. Sci.*, *57*(12), 5949–5949.

Venhuizen, F. G., van Ginneken, B., Liefers, B., van Asten, F., Schreur, V., Fauser, S., Hoyng, C., Theelen, T., & Sánchez, C. I. (2018). Deep learning approach for the detection and quantification of intraretinal cystoid fluid in multivendor optical coherence tomography. *Biomedical Optics Express*, *9*(4), 1545–1569. https://doi.org/10.1364/boe.9.001545

Wang, J., Zhang, M., Pechauer, A. D., Liu, L., Hwang, T. S., Wilson, D. J., Li, D., & Jia, Y. (2016). Automated volumetric segmentation of retinal fluid on optical coherence tomography. *Biomedical Optics Express*, *7*(4), 1577–1589. https://doi.org/10.1364/boe.7.001577

Wilkins, G. R., Houghton, O. M., & Oldenburg, A. L. (2012). Automated segmentation of intraretinal cystoid fluid in optical coherence tomography. *IEEE Transactions on Biomedical Engineering*, *59*(4), 1109–1114. https://doi.org/10.1109/TBME.2012.2184759

Wu, J., Philip, A. M., Podkowinski, D., Gerendas, B. S., Langs, G., Simader, C., Waldstein, S. M., & Schmidt-Erfurth, U. M. (2016). Multivendor Spectral-Domain Optical Coherence Tomography Dataset, Observer Annotation Performance Evaluation, and Standardized Evaluation Framework for Intraretinal Cystoid Fluid Segmentation. *Journal of Ophthalmology*, *2016*. https://doi.org/10.1155/2016/3898750

Yan, Q., Chen, B., Hu, Y., Cheng, J., Gong, Y., Yang, J., Liu, J., & Zhao, Y. (2020). Speckle reduction of OCT via super resolution reconstruction and its application on retinal layer segmentation. *Artificial Intelligence in Medicine*, *106*, 101871. https://doi.org/10.1016/j.artmed.2020.101871

Zhu, W., Zhang, L., Shi, F., Xiang, D., Wang, L., Guo, J., Yang, X., Chen, H., & Chen, X. (2017). Automated framework for intraretinal cystoid macular edema segmentation in three-dimensional optical coherence tomography images with macular hole. *Journal of Biomedical Optics*, *22*(7), 076014. https://doi.org/10.1117/1.jbo.22.7.076014